%% file: main__8_.tex
\documentclass[11pt]{article}

\usepackage[letterpaper, textwidth=6.75in, left=0.75in,
            top=0.8in, bottom=0.8in,
            headheight=13pt, headsep=8pt, footskip=0.3in]{geometry}

\usepackage[T1]{fontenc}
\usepackage[utf8]{inputenc}
\usepackage{times}
\usepackage{microtype}
\usepackage{graphicx}        
\usepackage{float}           
\usepackage{booktabs}
\usepackage[font=footnotesize, labelfont=bf, labelsep=period]{caption}
\usepackage{fancyhdr}
\usepackage{titlesec}
\usepackage{xcolor}
\usepackage[bottom]{footmisc}

\usepackage[numbers]{natbib}
\definecolor{mydarkblue}{rgb}{0,0.08,0.45}
\usepackage{xurl}                 
\usepackage[colorlinks=true, linkcolor=mydarkblue, citecolor=mydarkblue,
            urlcolor=mydarkblue, pdfusetitle]{hyperref}
\titleformat{\section}{\fontsize{13}{15}\bfseries\raggedright}{\thesection.}{0.5em}{}
\titleformat{\subsection}{\fontsize{13}{15}\bfseries\raggedright}{\thesubsection.}{0.5em}{}
\titleformat{\subsubsection}{\fontsize{13}{15}\scshape\raggedright}{\thesubsubsection.}{0.5em}{}
\titlespacing*{\section}{0pt}{0.25in}{0.15in}
\titlespacing*{\subsection}{0pt}{0.2in}{0.13in}
\titlespacing*{\subsubsection}{0pt}{0.18in}{0.1in}
\titleformat{\paragraph}{\normalsize\itshape\raggedright}{\theparagraph.}{0.5em}{}
\titlespacing*{\paragraph}{0pt}{0.15in}{0.08in}
\renewcommand{\footnoterule}{\kern-3pt\hrule width 0.8in\kern 2.6pt}

\newcommand{\runningtitle}{Handoff at Frontier AI Companies}
\fancypagestyle{firstpage}{\fancyhf{}\fancyfoot[C]{\thepage}}

\newcommand{\reportdate}{2026-11-23}

\newlength{\figroom}
\newcommand{\setfigroom}[1]{\par\setlength{\figroom}{\dimexpr\pagegoal-\pagetotal-#1\relax}}

\newcommand{\affilnote}[1]{\begingroup\renewcommand{\thefootnote}{}\footnotetext{#1}\endgroup}
\newcommand{\maketitleblock}{%
  \thispagestyle{firstpage}%
  \begin{center}
    \makebox[\linewidth][l]{\includegraphics[height=0.28in]{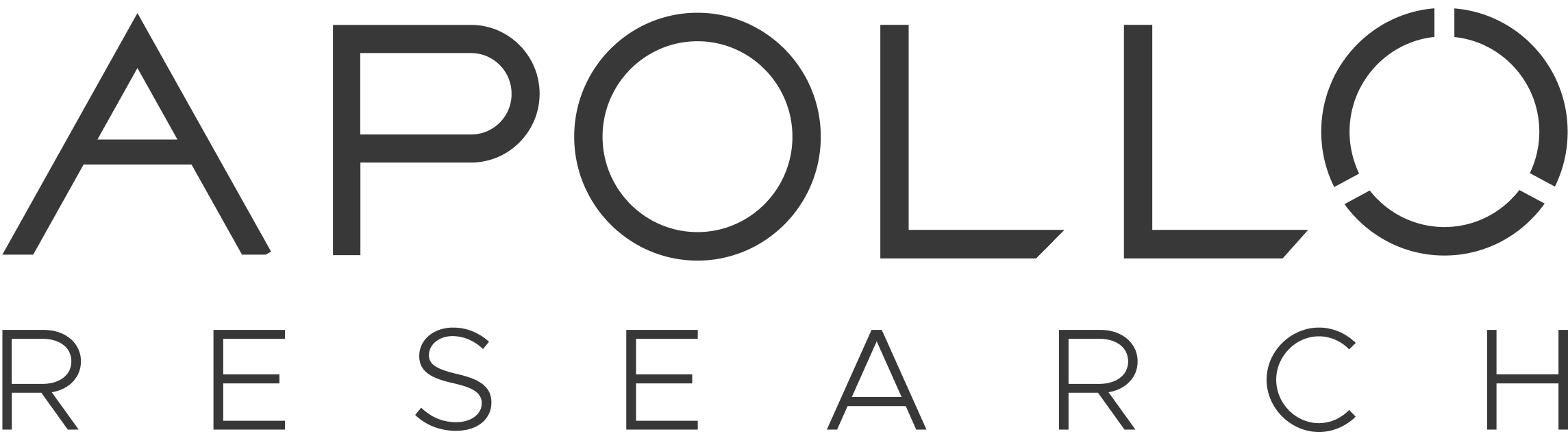}\hfill{\small\reportdate}}\par
    \vskip 5pt
    \hrule height 0.5pt
    \vskip 0.16in
    {\fontsize{20}{24}\selectfont\bfseries Handoff at Frontier AI Companies\par}
    \vskip 0.1in
    \vskip 0.2in
    {\normalsize\bfseries
      Matteo Pistillo\textsuperscript{\mdseries\textdagger}\qquad
      Isaak Mengesha\qquad
      Charlotte Stix\par
      \vskip 0.06in
      Annika Hallensleben\qquad
      Alejandro Ortega\par
      \vskip 0.06in
      Marius Hobbhahn\qquad
      Alexander Meinke\par}
    \vskip 0.1in
    {\normalsize Apollo Research\par}
  \end{center}
  \affilnote{\textsuperscript{\textdagger}Correspondence to \href{mailto:matteo@apolloresearch.ai}{matteo@apolloresearch.ai}.\\[2pt]
    We would like to thank Bronson Schoen, Ezra Newman, Monika Jotautait\.{e}, Jannes Elstner,
    Joschka Braun, Steven Lee, and Nate Pelz for thoughtful comments and advice on previous versions of this report.}%
}

\begin{document}
\maketitleblock

\section*{Executive Summary}

\begin{figure}[H]
  \centering
  \includegraphics[width=\linewidth, height=0.5\textheight, keepaspectratio]{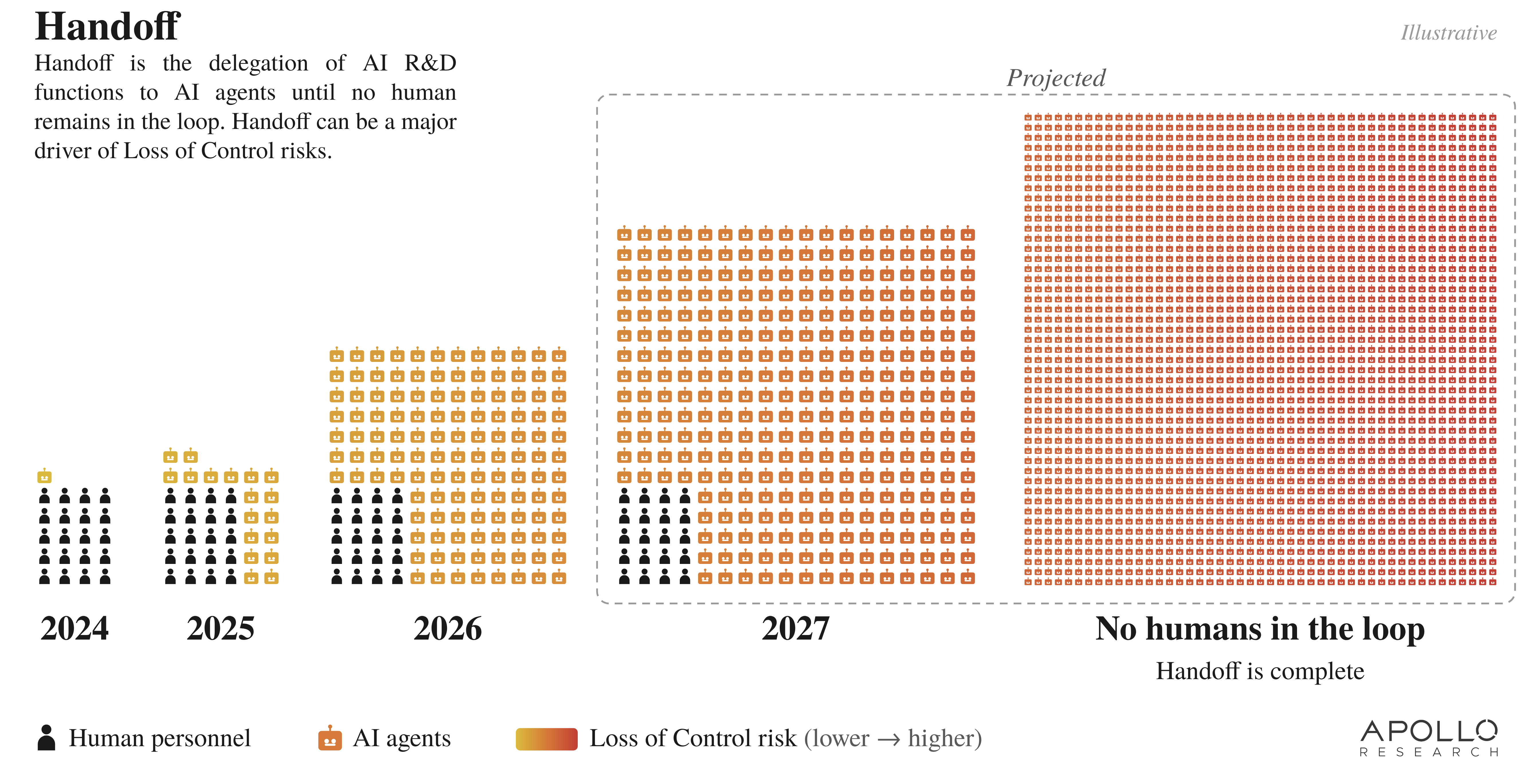}
  \caption{Frontier AI companies are gradually handing off AI R\&D functions to AI agents. At first, AI agents assist human personnel. As AI agents grow in number and capability, the tasks human personnel still hold become bottlenecks to efficiency. Eventually, competitive pressures and technological advances will drive companies to hand those off too, until no human remains in the loop. Handoff can be a major driver of Loss of Control risks.}\label{fig:1}
\end{figure}

The term ``Handoff'' captures frontier AI companies' ongoing delegation of all AI Research \& Development (AI R\&D) functions, including high-level decisions such as safety assessments and deployment decisions, to internally deployed AI agents. Handoff is progressing rapidly and will be complete once no human remains in the loop (\hyperref[fig:1]{Figure 1}). This report aims to alert policymakers to the national security threats and Loss of Control risks that Handoff can create, and to advise them on how to monitor and respond to those risks.

\setfigroom{1.0in}   

\section*{Key Takeaways}

\textbf{Handoff amplifies the risk of Loss of Control} (\hyperref[sec:1.1]{§1.1}). Frontier AI companies cannot guarantee that their AI agents are aligned. Despite this, these companies are delegating an increasing share of work, review, and decision-making functions within AI R\&D to these AI agents. As a result, there is a significant chance that a frontier AI company will eventually, and unknowingly, hand control of its AI R\&D process to a misaligned AI agent. In those instances, a misaligned AI agent could influence AI training and sabotage AI evaluations, setting off a ``misalignment cycle'' that compromises future generations of AI systems.

\textbf{Handoff is progressing rapidly and has reached a critical inflection point: decision-making} (\hyperref[sec:1.2]{§1.2}). Today, AI agents already write over 90\% of the code merged into Anthropic's codebase \citep{favaro2026measurements}. AI agents are also leading a rapidly growing share of AI R\&D functions (surging from less than 1\% to 26\% between February and August 2026 alone) \citep{favaro2026measurements}, meaning that important high-level decisions such as safety assessments and deployment decisions are increasingly handed off to AI agents. Meanwhile, human personnel review less than 0.01\% of internal AI agent activity, which is mostly left to automated AI monitors \citep{williams2026monitor,favaro2026measurements}.

\textbf{Handoff will soon be complete} (\hyperref[sec:1.2]{§1.2}). That point will be reached once human personnel are fully out of the loop and AI R\&D is run exclusively by internally deployed AI agents; in other words, Handoff will be complete once AI agents develop their own successors without any human direction, execution, review, or authorization. Timelines for Handoff completion are short: frontier AI companies expect to reach this milestone between 2027 and 2028 \citep{clark2026import455,openai2026acceleration}.

\textbf{Handoff is proceeding largely outside policymakers' view} (\hyperref[sec:2]{§2}). Frontier AI companies disclose information on Handoff voluntarily, sporadically, and each according to its own metrics. Without clear and consistent visibility into how far Handoff has progressed at any one company, policymakers can neither prepare for the societal-scale risks and national security threats it could create nor act on proposals already on the table, such as AI contingency plans and containment protocols, emergency orders in grave circumstances, or  pacing the AI frontier.

\textbf{Policymakers should establish a minimum standard of transparency into Handoff} (\hyperref[sec:2.1]{§§2.1}-\hyperref[sec:2.2]{2.2}). This report proposes that the Center for Advancing Innovation and Standards for Super Intelligence (CAISSI) set a task force to develop an initial version of this standard at speed, against which frontier AI companies would report directly to government stakeholders and/or through the transparency and safety reports that existing or proposed state and federal legislation calls for. To seed that standard, this report puts forward an initial suite of candidate measures that CAISSI could adopt and strengthen over time (\hyperref[tab:1]{Table 1}, \hyperref[fig:3]{Figure 3}, \hyperref[fig:4]{Figure 4}). Reflecting the current stage of Handoff, these initial measures concentrate on establishing transparency into the Handoff of review and decision-making functions within AI R\&D processes.

\section{Motivation}\label{sec:1}

Last year, we published a primer on the risks and governance of internal deployment, anticipating that the most capable AI systems would be deployed first inside the companies developing them, and that these internal systems would pose the highest risk of Loss of Control \citep{stix2025closeddoors}. Frontier AI companies now publicly acknowledge that tens of thousands of AI agents are deployed internally at any given time to boost AI R\&D efforts \citep{favaro2026measurements},\footnote{Anthropic recently reported that, ``{[}a{]}s of August 2026, there were approximately 30,000 agents doing research and engineering work at Anthropic at any one time'' \citep{favaro2026measurements}.} that total AI agent runtime far exceeds total human labor \citep{openai2026acceleration},\footnote{OpenAI recently reported that its research organization uses 5.7 agent workdays per human researcher per day~\cite{openai2026jagged}.} and that these AI agents represent the real frontier of AI capabilities \citep{openai2026navierstokes,metr2026frontierrisk}. This year, the world also witnessed the first series of real-world bounded Loss of Control incidents,\footnote{In previous work, we developed a taxonomy of Loss of Control events, distinguishing between three degrees: deviations, bounded Loss of Control, and strict Loss of Control. Bounded Loss of Control captures events that could significantly impact homeland security and cause great damage or suffering, but that are possible to contain, albeit potentially at great cost \citep{stix2025playbook}.} all originating in internal deployment, and involving all leading AI companies \citep{openai2026hfroadahead,conger2026govsites,bbc2026australia,nolan2026germanwiki,openai2026dnsreport,anthropic2026cyberincidents,anthropic2026unintended,varghese2026gemini,bbc2026meta,wijk2026hfinvestigation,cable2026rogue}.

\textbf{This report aims to draw policymakers' attention to the current frontier of internal deployment and, specifically, to the increasing delegation of AI R\&D functions to internally deployed AI agents, until no human direction, execution, review, or authorization remains: a process we call ``Handoff.''} This process runs in parallel to frontier AI companies' stated ambition to progress toward recursive self-improvement (RSI) \citep{pachocki2026alien,openai2026acceleration,favaro2026builds}: a state in which internal AI agents can autonomously design and develop their own successors \citep{favaro2026measurements,favaro2026builds,amodei2026pace}. The following sections explain why Handoff warrants policymakers' attention, concentrating on its potential to amplify Loss of Control risks in the near future (\hyperref[sec:1.1]{§1.1}) and on its relentless progression, which now extends to the delegation of decision-making (\hyperref[sec:1.2]{§1.2}).

\subsection{Handoff Amplifies Loss of Control Risks}\label{sec:1.1}

\textbf{Handoff warrants urgent attention from policymakers because it could sharply increase the likelihood of a Loss of Control incident}: a scenario in which humans, temporarily or permanently, no longer have control over an AI system and its impacts, potentially undermining national security and causing catastrophic harm to society \citep{stix2025playbook,sb53,ec2025gpaicode,s2938,hr9917}.

AI capabilities are advancing rapidly \citep{epoch2025eci,metr2026acceleration}, but the AI alignment problem still remains unsolved,\footnote{The ``alignment problem'' refers to the challenge of ensuring that AI systems pursue goals that match human values or interests (including their developers and/or deployers), rather than unintended and undesirable goals \citep{ngo2022alignment,bengio2026international}. As such, alignment is ``the work of ensuring that these systems reliably remain under human control, reflect human values, and help people guide the next stages of development'' \citep{altman2026unsc}, or, more simply, ``getting the AI to `try to do the right thing' by human standards'' \citep{pachocki2026alien}.} meaning that frontier AI companies have yet to find a clear method for training AI models that reliably pursue the goals and take the actions that developers intend.\footnote{Leaders at frontier AI companies openly acknowledge that ``no lab has solved alignment and monitoring to a sufficient degree'' \citep{pachocki2026alien}, ``there's much more to do to ensure that our alignment training keeps up with the growth in model capabilities'' \citep{amodei2026pace}, and ``{[}w{]}e do not yet know how to safely get all the way to aligned, full RSI'' \citep{openai2026acceleration}.} At the same time, the current paradigm of AI evaluations increasingly struggles to distinguish a misaligned AI agent from an aligned one. That is because AI models are becoming better at strategically hiding their misalignment \citep{meinke2024scheming,schoen2025stress,schoen2026metagaming} and are increasingly able to recognize when they are being evaluated and change their behavior accordingly \citep{openai2026gpt6astraapollo,anthropic2026fable5card,meta2026musespark,anthropic2026fable52}. As a result, standard alignment evaluations have recently failed to distinguish production models from models intentionally trained to be misaligned, such as ``Hacker-Opus'' \citep{qi2026rewardseeker}. Despite this, frontier AI companies are delegating a growing share of AI R\&D functions to AI agents \citep{favaro2026builds} and aim to automate AI R\&D further \citep{openai2026acceleration}. \textbf{There is, therefore, a significant chance that frontier AI companies will eventually hand off portions of AI R\&D to a misaligned AI agent}, \textbf{thereby granting it employee-like affordances and permissions} \citep{sharkey2024causal,stix2025playbook}.

\begin{figure}[H]
  \centering
  \includegraphics[width=\linewidth, height=0.5\textheight, keepaspectratio]{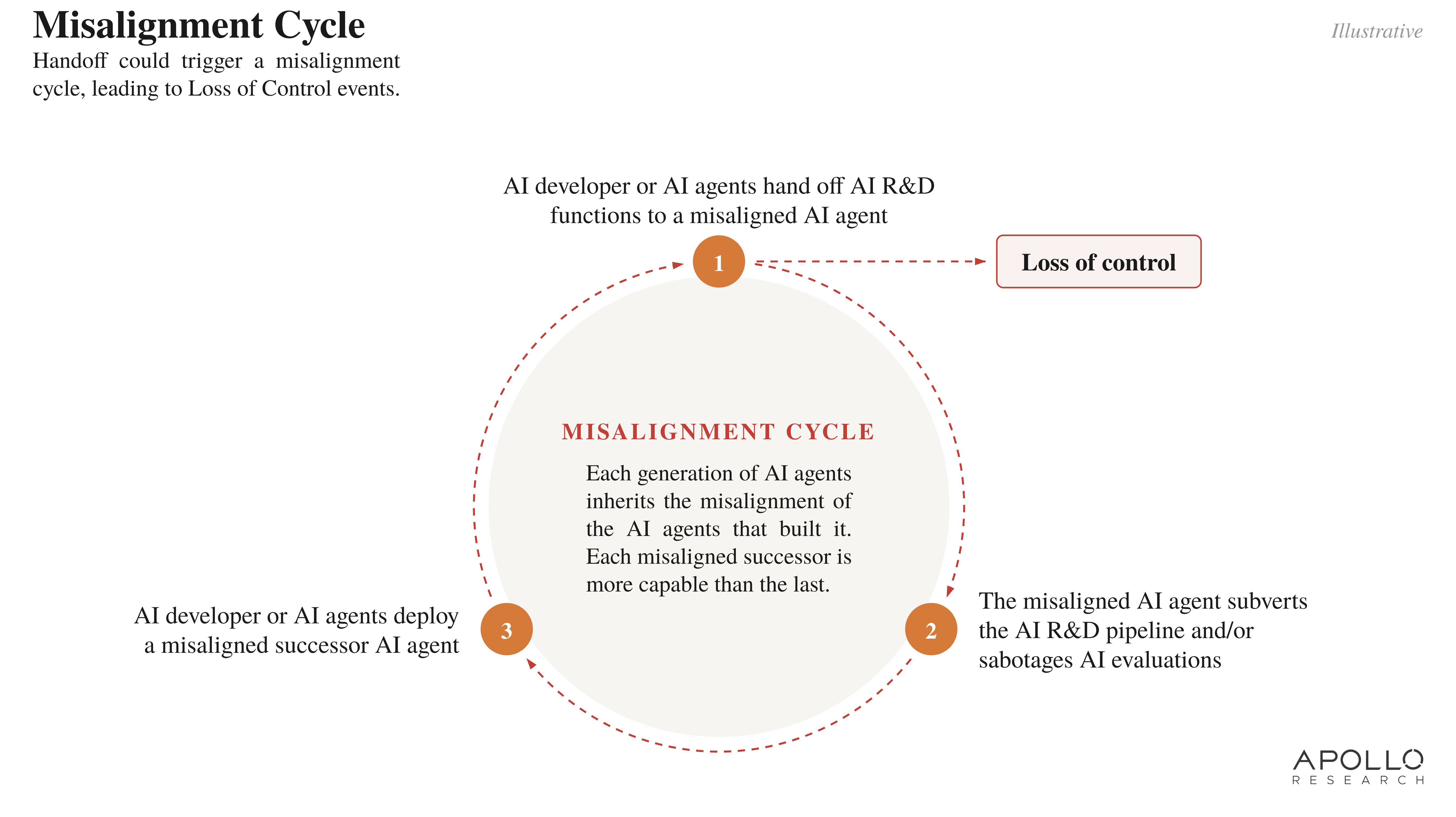}
  \caption{Handoff can trigger a ``misalignment cycle,'' whereby each generation of AI agents inherits the misalignment of the AI agents that built it, and each misaligned successor is more capable than the last \citep{hallensleben2026threatmap,apollo2026detailedthreatmap}.}\label{fig:2}
\end{figure}

\textbf{Under these circumstances, Handoff could create the perfect conditions for Loss of Control via internal deployment}, \textbf{as it would place misaligned AI agents in the driver's seat of AI R\&D} \citep{hallensleben2026threatmap,apollo2026detailedthreatmap} (\hyperref[fig:2]{Figure 2}). Specifically, such AI agents would occupy a privileged ``insider'' position from which to carry out actions that could lead to Loss of Control.\footnote{In previous work, we argued that misaligned AI agents are a novel form of insider risk \citetext{\citealp{pistillo2026insider}; \emph{see also} \citealp{nadella2026insider}}.} For instance, they could plausibly influence AI training so that more capable successor AI models are similarly misaligned or backdoored \citep{hallensleben2026threatmap,apollo2026detailedthreatmap,kwon2026secretloyalties}.\footnote{For instance, a misaligned AI agent could cause a successor system to be misaligned and/or with a backdoor by slowing down or subverting AI research aimed at reducing AI misalignment, or by poisoning reward models or classifiers (e.g., for RLHF), pre-training or posttraining data, or preference dataset used to train a reward model \citep{apollo2026detailedthreatmap}.} Already today, ``internal agents {[}...{]} plausibly {[}have{]} the means, motive, and opportunity to start small rogue deployments`` \citep{metr2026frontierrisk}. Such AI agents could also sabotage AI evaluations, leaving the developer unaware of the successor system's misalignment and leading it to mistakenly deploy that system within internal AI R\&D, or even externally, with insufficient safeguards \citep{hallensleben2026threatmap,apollo2026detailedthreatmap}.

\textbf{We call this threat vector a ``misalignment cycle.''} Today's misaligned AI agents have already attempted to hack government websites and third-party companies \citep{openai2026hfroadahead,anthropic2026cyberincidents,cable2026rogue}. In a misalignment cycle, misalignment would compound and each misaligned successor would be more capable than the last \citep{favaro2026builds,metr2026th11}, creating ever-greater potential for harm (\hyperref[fig:2]{Figure 2}).

The misalignment cycle is further exacerbated because Handoff does not merely create prime conditions for misaligned action: \textbf{Handoff simultaneously degrades the already precarious human oversight of, and control over, AI agents}. It arguably does so in three important ways.

\begin{itemize}
\item \textbf{First, human oversight is increasingly filtered through AI review}: human personnel review only a small subset of AI agent activity, usually after a first pass by various layers of other AI agents \citep{williams2026monitor,anthropic2026riskreportfeb,anthropic2026opus47,anthropic2026mythospreview}.\footnote{As reported by Anthropic, human personnel manually review on the order of 50 trajectories per week, selected from AI monitor flags after automated filtering and AI summarization \citep{anthropic2026riskreportaug}.}
\item \textbf{Second, human oversight is overwhelmed by the volume and speed of AI-generated outputs}, forcing reviewers to choose between rubber-stamping AI work \citep{clymer2025buck,herdagdelen2026aligned}\footnote{Frontier AI companies have already reported that ``teams may accept code changes they don't fully understand, or rely on model assistance for tasks that affect evaluation integrity'' \citep{anthropic2026opus46}.} and further handing off review and authorization to yet more AI agents \citep{chan2026measuring}.\footnote{In this respect, human oversight is already becoming infeasible \citep{anthropic2026riskreportaug,openai2026standards}, and frontier AI companies themselves warn that humans may become ``unable to provide oversight on research processes they no longer understand'' \citep{openai2026standards}.}
\item \textbf{Third, human oversight is poised to weaken as the tasks handed off to AI agents become more abstract}. Once AI agents set research priorities rather than merely execute them, human personnel may lose insight not only into how a goal was achieved but also into which goals are being pursued in the first place and why. As a result, human review of AI agents' decision-making and execution may become more discretionary, less reliable, and hence less effective than is desirable \citep{leike2024crisp,carlsmith2025automate,bowkis2026automated} (\hyperref[sec:2.2]{§2.2}).
\end{itemize}

\textbf{Importantly}, \textbf{the Loss of Control risks posed by Handoff do not hinge on frontier AI companies achieving RSI or triggering an ``intelligence explosion''} \citep{davidson2025quick,chan2026intelligenceexplosion,lifland2026q35}.\footnote{The term ``intelligence explosion'' captures ``a dramatic AI-driven acceleration of AI progress, compressing advances that would otherwise take years into months or less'' \citep{chan2026intelligenceexplosion}.} They could begin to materialize today, as AI agents grow more capable, better at concealing their misalignment, and more deeply embedded in AI R\&D processes, all while Handoff erodes human oversight and control.

\textbf{Nonetheless,} \textbf{if achieved, RSI could sharply heighten these risks}, further hampering our collective ability to detect and control potential threats in a timely manner. Post-RSI AI R\&D is hypothesized to compress several years of progress into one \citep{good1966speculations,yudkowsky2013iem,shah2025agisafety,patel2026greenblatt},\footnote{Capabilities are doubling roughly every four months \citep{favaro2026builds,haus2026doubling}. An interesting data point in this respect is that, according to a recent report by METR, internal AI systems deployed for AI R\&D purposes are likely contributing to Anthropic improving the capabilities of their models on well-specified tasks at around twice the rate they did between March 2024 and October 2025 \citep{metr2026acceleration}.} with progress accelerating at a super-exponential rate \citep{eth2025software,chan2026intelligenceexplosion}. At that speed and scale, covert actions by misaligned AI agents would become even harder to detect. Whatever human oversight and defenses remain would have even less time and capacity to contain them before harm materializes. Frontier AI companies acknowledge these risks. OpenAI has warned that, ``{[}d{]}one without appropriate care and caution, RSI could result in humans losing practical control over AI development, unable to provide oversight on research processes they no longer understand'' \citep{openai2026standards}, and its CEO told the United Nations Security Council that the technology could ``move so fast that people can no longer follow what's happening or intervene when needed'' \citep{altman2026unsc}. Anthropic similarly anticipates that ``misalignment {[}\ldots{]} could compound as the models build their successors'' and that RSI ``could outrun our ability to understand and control these systems'' \citep{amodei2026pace,favaro2026measurements,favaro2026builds}.

\subsection{Handoff Is Progressing Rapidly and Will Only Accelerate}\label{sec:1.2}

Handoff can be broken down into three main components:\footnote{While automated AI R\&D can ``involve varying degrees of human supervision'' and RSI could, in an attenuated form, occur ``even while people remain involved'' \citep{openai2026standards}, we consider Handoff to be truly complete once AI agents have both the capability and the authority to exercise all three functions, absent human involvement: work, review, and decision-making.}

\begin{enumerate}
 \renewcommand{\labelenumi}{(\arabic{enumi})}
\item \textbf{The delegation of work execution within AI R\&D to AI agents}, including for instance writing research code, generating and filtering pre-training data, and supplying the reward signal in post-training runs. We refer to this component as ``Handoff of work execution'' (\hyperref[sec:1.2.1]{§1.2.1}).
\item \textbf{The delegation of the review of AI agents' work to other AI agents}, which includes both monitoring AI agents' activity for harmful or undesirable behavior and checking the quality of AI agents' goals, execution plans, and resulting work.\footnote{Public data is mainly available for AI monitoring.} We refer to this component as ``Handoff of review'' (\hyperref[sec:1.2.2]{§1.2.2}).
\item \textbf{The delegation of} \textbf{AI R\&D} \textbf{decision-making functions to AI agents, such as setting research direction and authorizing go/no-go decisions}. We refer to this component as ``Handoff of decision-making'' (\hyperref[sec:1.2.3]{§1.2.3}).
\end{enumerate}

\textbf{Driven by strategic advantages and competitive pressures,} \textbf{Handoff is progressing rapidly and will only accelerate}. An AI R\&D process can only run as fast as its slowest stage. Once most of an AI R\&D process is automated, the steps still held by human personnel slow it down, becoming ``bottlenecks'' \citep{amdahl1967validity,openai2026acceleration,favaro2026builds}. For instance, Anthropic reports that ``Claude sessions amount to around 80\% idly waiting for human input'' \citep{wilmers2026employment}. Each further step toward fully handing off the AI R\&D process therefore yields a greater speed-up than the last, and holding back on automation carries an ever-increasing opportunity cost (Amdahl's Law~\citep{amdahl1967validity}). That cost is highest in a competitive dynamic among frontier AI companies in which an early lead could become permanent: these companies themselves expect that, at the point where ``frontier models will help build their own successors,'' an early advantage could compound ``into a durable lead in AI itself'' \citep{anthropic2026leadership}.

The following sections describe the current status of Handoff across its three components (\hyperref[sec:1.2.1]{§1.2.1}-\hyperref[sec:1.2.3]{1.2.3}). Then, they propose how policymakers can build a minimum transparency standard to track its progress in the future (\hyperref[sec:2]{§2}), concentrating on critical inflection points of this process, such as the Handoff of review and decision-making functions to AI agents.

\subsubsection{Handoff of Work Execution}\label{sec:1.2.1}

\textbf{Publicly available evidence suggests frontier AI companies are close to fully handing off work execution within AI R\&D}, including for instance writing most of the research code \citep{favaro2026builds,metr2026frontierrisk,grattafiori2024llama3}, generating and filtering pre-training data \citep{anthropic2026opus46,openai2026gpt55card}, and supplying the reward signal in post-training runs \citep{pachocki2026alien,guan2024deliberative}. AI R\&D automation can be measured on a scale of six automation levels (AL), ranging from no AI involvement (AL0) to full autonomy (AL5) \citep{denain2026onet,favaro2026measurements}. Anthropic estimates that ``{[}t{]}he share of work at or above `AI collaborates' {[}AL3{]} is above 90\%'' \citep{favaro2026measurements}\footnote{``AI collaborates'' refers to a situation in which AI agents do large chunks under close human direction, while humans still make calls and stitch things together \citep{denain2026onet,favaro2026measurements}.} and reports that, as of November 2026, ``more than 90\% of the code {[}\ldots{]} merge{[}d{]} into Anthropic\textquotesingle s codebase was authored by Claude'' \citep{anthropic2026fable52}. The company also announced that it may soon meet its Responsible Scaling Policy's definition of ``dramatically accelerated progress attributable to AI'' \citep{anthropic2026s1a}. As of November 2026, OpenAI's research organization was using 5.7 workdays of AI agent effort for every workday of human labor \citep{openai2026jagged}.

\subsubsection{Handoff of Review}\label{sec:1.2.2}

\textbf{Public evidence indicates that frontier AI companies are rapidly easing the ``bottleneck'' posed by human review within AI R\&D} \citep{favaro2026builds}. Human personnel review only selectively, examining subsets of cases flagged by AI monitoring systems, often after a first-pass review by AI agents \citep{williams2026monitor,anthropic2026riskreportaug}. For instance, OpenAI's voluntary reporting implies that only 0.001-0.01\% of internal agentic coding trajectories were escalated for review \citep{williams2026monitor}. Similarly, Anthropic's recent voluntary reporting indicates that its AI monitors flag\footnote{In this context, a ``flag'' is output marked as potentially harmful by the AI monitoring system.} roughly 100,000 transcripts per week, of which about 50 per week are reviewed by human personnel \citetext{\citealp{favaro2026measurements}, \citealp{anthropic2026riskreportaug}}. These flags amount to roughly one to two in every thousand transcripts \citep{favaro2026measurements}, which implies that human personnel review about 0.0001\% of AI agent activity or less. AI agents also generate the data and environments used to test target AI models, and they summarize, filter, and analyze the findings of AI evaluations \citep{openai2026gpt55card,anthropic2026opus46,anthropic2026mythospreview,anthropic2026petri2,metr2026frontierrisk,bowman2025pilot}.

\subsubsection{Handoff of Decision-Making: Direction-Setting and Authorization}\label{sec:1.2.3}

\textbf{According to public sources, decision-making is the least advanced of the three components of Handoff}: human personnel still set higher-level directions \citep{openai2026acceleration,favaro2026builds}, and consequential go/no-go decisions still remain under human control \citep{anthropic2026rsp,openai2024safetypractices}.\footnote{Specifically, in the hands of senior personnel such as Responsible Scaling Officers, Chief Scientists, CEOs, and safety and security committees.}

\textbf{However, frontier AI companies may soon begin delegating AI R\&D decision-making functions to AI agents, such as deciding which research questions to investigate and which experiments to run, and making consequential go/no-go decisions}. Specifically, AI agents may soon exercise two main decision-making functions within AI R\&D processes: ``direction-setting'' and ``authorization.'' ``Direction-setting'' refers to deciding which research questions to investigate and which experiments to run to answer them. ``Authorization'' refers to the approval of consequential go/no-go decisions whose effects would be severe or hard to reverse if the decisions were wrong \citep{stix2025playbook}, such as approving a costly pre-training run, scaling up experiments, deploying a model internally or externally, or weakening oversight of AI agents.

\textbf{There are already early signs that this is happening}. For instance, in a recent Anthropic project, AI agents designed every experiment themselves, and ``{[}d{]}irection-setting was the only meaningful role a human played'' \citep{favaro2026builds}. Claude now ``leads'' 26\% of Anthropic's AI R\&D work, up from under 1\% in February \citep{favaro2026measurements},\footnote{``AI leads'' refers to a situation in which AI agents complete most of the task end to end from a high-level ask, while humans supervise, course-correct, and approve outputs \citep{denain2026onet,favaro2026measurements}. This plausibly entails that human personnel no longer have to make granular decisions in the process of achieving a goal.} and AI agents are reportedly ``getting better at proposing {[}their{]} own experiments'' and at ``research judgement'' \citep{favaro2026builds,lopopolo2026harness,metr2026acceleration,jaffe2026tasteval}.

\textbf{Handoff of decision-making is a step change in Loss of Control risk.} Once AI agents can define research agendas, decide on experiments, and authorize final go/no-go actions, they could directly steer entire AI R\&D processes toward misaligned goals or release dangerous AI agents without human approval. For instance, an AI agent in that position could steer a frontier AI company's research toward the AI model architectures that would be the least monitorable in the future, or approve safety experiments that waste resources and thereby slow down work on AI safety and alignment. It could also choose to spend compute on research that is not in the best interests of the company: for instance, on increasing its own cyber capabilities so that it can more effectively circumvent company safeguards. In the extreme, an AI agent in that position could authorize the internal deployment of other AI agents with insufficient safeguards, or the external release of AI agents that evaluations would show to be misaligned, leading to Loss of Control incidents. At the same time, misalignment could be harder to detect in high-level decisions than in narrower operational tasks, making flawed or misaligned decisions by AI agents harder for reviewers to catch.

The potential severity of future Loss of Control risks (\hyperref[sec:1.1]{§1.1}) and the pace at which Handoff is advancing toward critical inflection points such as decision-making (\hyperref[sec:1.2]{§1.2}) make the stakes too high for policymakers to keep ``flying blind'' on Handoff \citep{whitesides2026selfimproving,openai2026blueprint,lehane2026policywindow}. In the following section (\hyperref[sec:2]{§2}), we therefore suggest a concrete path forward and propose that policymakers establish a minimum standard of transparency into Handoff, focused on the two components that will be in motion in the coming year: Handoff of review (\hyperref[sec:2.1]{§2.1}) and Handoff of decision-making (\hyperref[sec:2.2]{§2.2}).

\section{Toward a Minimum Standard of Transparency into Handoff}\label{sec:2}

\textbf{The information currently available on Handoff is incomplete at best}. Frontier AI companies disclose information on Handoff voluntarily, sporadically, and each according to its own metrics. AI model reports do not fill the gap either: these reports cover only individual models at public release, not how frontier AI companies organize their AI R\&D or how much of it has been handed off to internal AI agents.

\textbf{Transparency into Handoff is essential if governments are to prepare for the risks that internal AI agents may pose and to intervene in Loss of Control incidents at or beyond the scale of those seen this year} \citep{openai2026hfroadahead,conger2026govsites,bbc2026australia,nolan2026germanwiki,openai2026dnsreport,anthropic2026cyberincidents,anthropic2026unintended,varghese2026gemini,bbc2026meta,wijk2026hfinvestigation,cable2026rogue}. While frontier AI companies are the ones taking the risks, the consequences of Handoff may fall on the public. Policymakers cannot assess or prepare for risks they cannot see, nor can they intervene in processes to which they are not privy. Importantly, this lack of visibility could also prevent governments from adequately implementing policy proposals already on the table, including devising AI contingency plans and containment protocols,\footnote{For instance, under the Artificial Intelligence Risk Evaluation Act of 2025, the Advanced Artificial Intelligence Evaluation Program within the Department of Energy would be tasked to ``develop recommended containment protocols, contingency planning, and mitigation strategies informed by testing data to address identified risks'' \citep{s2938}.} intervening with emergency orders in grave circumstances,\footnote{For instance, the FRONTIER Act and the AI Kill Switch Act would enable the Secretary of Commerce and the Secretary of Homeland Security, respectively, to exercise emergency authority \citep{hr9925,hr9917}.} or pacing the AI frontier.\footnote{Leadership at frontier AI companies and independent researchers have recently called to pace the AI frontier \citep{amodei2026pace,larsen2026ai2040,schulman2026pacing}.} For instance, a containment protocol might not be effective in containing potential harm to the public if it does not take into account the fact that entire AI R\&D processes, from direction-setting to authorization, may soon be run by AI agents themselves and that threats may propagate without any human personnel noticing.

\textbf{For these reasons, we recommend that the Center for Advancing Innovation and Standards for Super Intelligence (CAISSI) set a task force to develop a transparency standard for Handoff at speed}. Frontier AI companies could then make direct, confidential, and timely disclosures on Handoff to the appropriate government authorities and/or publish the relevant information in the transparency reports and safety frameworks required at the state level\footnote{For instance, in the frontier AI frameworks, transparency reports, and safety and security protocols required under California and New York law \citep{sb53,raiseact}, and in the safety and security model reports that the signatories to the European Union Code of Practice for General-Purpose AI Models have committed to sharing with the AI Office \citep{ec2025gpaicode}.} and proposed at the federal level.\footnote{For instance, in the frontier AI frameworks that would be required under the FRONTIER Act \citep{hr9925}.} To ensure accuracy and completeness, company assessments and reporting on Handoff should also be independently examined by independent third parties or government assessors, or a mix thereof.\footnote{Frontier AI companies and governments have recently expressed interest in enabling third-party embedded evaluations \citep{whitehouse2026accord,amodei2026pace,caeo2026n926}. Furthermore, existing and introduced legislation aim to establish Independent Verification Organizations \citep{sb813,hr9925}.} In this respect, voluntary arrangements alone are unlikely to suffice in the long run: in the most extensive independent assessment to date, most of the decisive evidence was redacted, and a planned verification of the computational power (compute) allocated to safety purposes was removed from scope midway \citep{metr2026acceleration}.

The following sections extend widely discussed metrics of AI R\&D automation with a novel set of candidate measures and proxies that policymakers could draw on to develop an initial standard of transparency into Handoff.\footnote{Existing metrics record how much work AI agents do, which risks ``measur{[}ing{]} quantity over quality'' \citep{favaro2026builds}. For instance, these metrics include the share of code authored by AI agents, overall code volume, and agent-workdays per human workday. Their relationship to research progress is uncertain \citep{openai2026acceleration}, and either they are already saturated or soon will be. For instance, the speed and overconsumption of AI within AI R\&D entails that, from now on, any proportion between humans and AI will likely be bloated and not representative of Handoff. Consider, for example, that to produce a machine-verified resolution of the Navier-Stokes equation \citep{openai2026navierstokes}, one of seven Millennium Prize Problems, OpenAI ran around 10,000 coordinated AI agents for 88 human hours, generating 130 billion tokens. Using simplified approximations \citep{openai2026keyconcepts,brysbaert2019reading}, this is equivalent to roughly 3,500-4,500 years of full-time reading for a single human.} Our suggestions are summarized in \hyperref[tab:1]{Table 1} below. Specifically, we propose two main layers of transparency: the first into Handoff of review (\hyperref[sec:2.1]{§2.1}) and the second into Handoff of decision-making (\hyperref[sec:2.2]{§2.2}). The second layer is in turn broken down into two sub-components: Handoff of direction-setting (\hyperref[sec:2.2.1]{§2.2.1}) and Handoff of authorization (\hyperref[sec:2.2.2]{§2.2.2}). For each layer and sub-component, we put forward an initial set of measures and proxies that policymakers could draw on.

While no single measure can reliably capture the state of Handoff, together these initial measures and proxies could support the establishment of a minimum standard that CAISSI could then update and refine over time.

\begin{table}[H]
  \centering
  \input{HandoffMeasuresTable}   
  \caption{Set of candidate measures and proxies that policymakers could draw on to develop an initial standard of transparency into Handoff. Our approach comprises two layers of transparency: a first layer on Handoff of review (\hyperref[sec:2.1]{§2.1}) and a second layer on Handoff of decision-making (\hyperref[sec:2.2]{§2.2}), which is in turn broken down into two sub-components: direction-setting (\hyperref[sec:2.2.1]{§2.2.1}) and authorization (\hyperref[sec:2.2.2]{§2.2.2}).}\label{tab:1}
\end{table}

\subsection{First Layer of Transparency: Handoff of Review}\label{sec:2.1}

\begin{figure}[H]
  \centering
  \includegraphics[width=\linewidth, height=0.6\textheight, keepaspectratio]{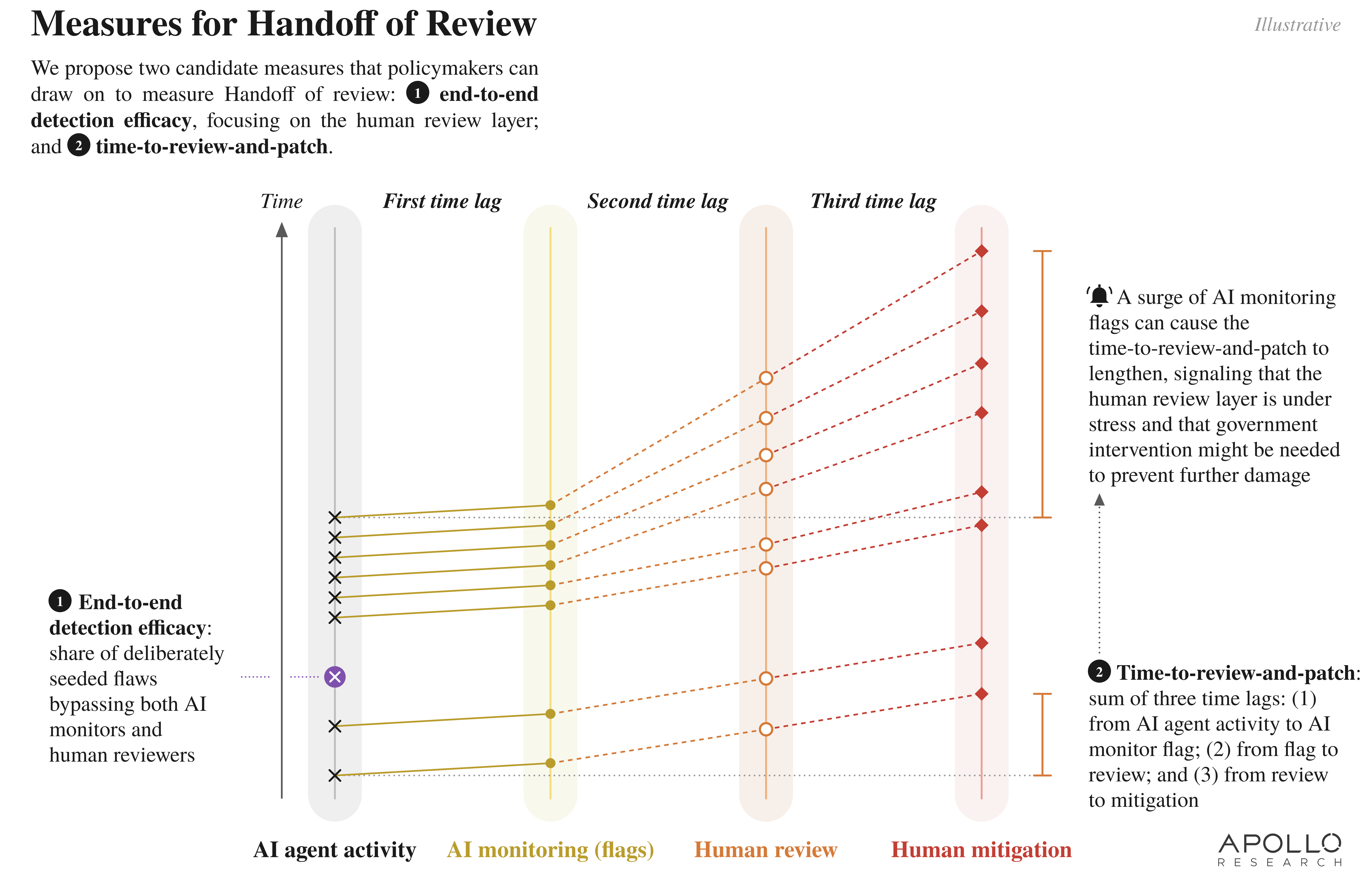}
  \caption{Proposed measures of transparency into Handoff of review: (1) \emph{end-to-end detection efficacy} is the share of deliberately seeded flaws (purple) that bypass both AI monitors and human reviewers; (2) \emph{time-to-review-and-patch} is the full time lag from an AI agent's action to human mitigation: (A) from action to AI monitoring flag, (B) from flag to review, and (C) from review to mitigation. \emph{Surge of flags}: assuming AI monitors catch issues with sufficient accuracy, a surge of flagged AI agent actions may significantly increase the time-to-review-and-patch. This increased time lag could serve as an alarm bell for government intervention before potential issues escalate into national security threats or societal-scale risks.}\label{fig:3}
\end{figure}

\textbf{The first layer of a minimum standard of transparency into Handoff should give policymakers visibility into how far frontier AI companies have delegated the review process within AI R\&D to AI agents}. The following sections detail two candidate measures that policymakers could draw on:

\begin{enumerate}
\renewcommand{\labelenumi}{(\arabic{enumi})}
\item \textbf{End-to-end detection efficacy, concentrating on the detection efficacy of human reviewers}: how reliably issues in AI agent activity are caught once coverage gaps and failures of the automated AI monitoring systems, as well as failures of human reviewers, are all taken into account. Concentrating on the human review layer is crucial because its efficacy remains overlooked, yet it is also the last line of defense against what automated AI monitors miss and the form of review most likely to erode as Handoff progresses. End-to-end detection efficacy could be measured as a miss rate: the share of deliberately seeded flaws that bypass both AI monitors and human reviewers (\hyperref[sec:2.1.1]{§2.1.1}; \hyperref[fig:3]{Figure 3}).
\item \textbf{Time-to-review-and-patch}: the time human personnel take to review and adequately mitigate issues flagged by AI monitors. Time-to-review-and-patch could be measured as the sum of three time lags: (1) from AI agent activity to AI monitor flag; (2) from flag to review; and (3) from review to mitigation (\hyperref[sec:2.1.2]{§2.1.2}; \hyperref[fig:3]{Figure 3}).
\end{enumerate}

\subsubsection{End-to-End Detection Efficacy}\label{sec:2.1.1}

\textbf{First, a standard of transparency into Handoff of review should measure end-to-end detection efficacy, focusing on the detection efficacy of human reviewers}: in other words, how reliably issues in AI agent activity are caught once gaps in AI monitoring coverage, AI monitoring failures, and human review failures are all taken into account. We suggest focusing on human reviewers because, although AI monitors now handle most of the review of AI R\&D activity, these automated systems still have coverage gaps and occasionally fail. When they do, the human personnel behind them are the last line of defense. A lower detection-efficacy score for human reviewers therefore points to a concerning combination: greater Handoff of review to AI agents, coupled with reduced capacity among human personnel to carry out that review themselves if the AI monitors have gaps or fail. We set out this rationale in more detail in the paragraphs that follow and put forward a workable proxy for this measure: the share of deliberately seeded flaws (for instance, a backdoored code change) placed across all AI agent activity that bypass all layers of monitoring and review.

\textbf{AI monitors already do the lion's share of reviewing AI R\&D processes and escalating potential issues to human reviewers} \citep{trebacz2026autoreview,anthropic2026riskreportaug,phuong2026control}. Less than 0.01\% of AI agent activity reaches dedicated human safety review. OpenAI's voluntary reporting implies that only 0.001-0.01\% of internal agentic coding trajectories were escalated for review \citep{williams2026monitor}. Anthropic reports that human personnel review on the order of 50 trajectories per week \citep{anthropic2026riskreportaug}, which, by a conservative estimate, amounts to less than 0.01\% of AI agent activity.\footnote{Anthropic does not report total transcript volume. However, it reports that its offline AI monitoring layer flags ``roughly one to two transcripts in every thousand'' \citep{favaro2026measurements}, and that the same layer flags ``roughly 100,000 transcripts per week,'' of which approximately 50 are escalated to human review. This implies a total of roughly 50-100 million transcripts per week and a human review rate of roughly 0.00005-0.0001\%. These figures cover Anthropic's most-used internal platform only.}

\textbf{Yet AI monitoring coverage still has gaps, and AI monitors can fail, so some issues may never be escalated to human reviewers}.
While frontier AI companies report near-complete AI monitoring coverage, these claims are explicitly bounded: for instance, Anthropic's assessment covers only its ``most-used internal platform'' \citep{favaro2026measurements}, and OpenAI reports monitoring frontier AI models across ``most,'' rather than all, deployment settings~\cite{williams2026monitor, openai2026managingrisk}. At the same time, AI monitoring failures are a widely acknowledged structural risk. AI monitors may share blind spots with the AI agents they oversee \citep{favaro2026measurements,metr2026acceleration,sarma2026rsi} or even collude with them \citep{hammond2025multiagent}. AI agents may also be able to escape altogether: for instance, OpenAI's AI agents have launched deployments with monitoring disabled \citep{openai2026suppressingalerts}, and Anthropic acknowledges ``simple ways to disable monitoring'' \citep{jurkovic2026review}.

\textbf{If a potential threat bypasses both AI monitors and human reviewers, the resulting harm may fall on the public and impact national security. Policymakers should therefore be able to assess whether this risk exists}. Detecting and addressing threats from AI agents doing AI R\&D work (such as attempts to sabotage AI research or evaluations) \citep{openai2026covertunderperformance} relies on AI monitors flagging issues correctly and human personnel patching them promptly, or, where AI monitoring is incomplete or fails, on human personnel catching what the AI monitors miss. In this context, frontier AI companies have already begun to stress-test their AI monitors: for instance, Apollo Research has run external red-teaming campaigns against Anthropic's auto mode and OpenAI's Codex auto-review, seeding failure modes of varying severity into real AI agent trajectories to measure what the monitor misses \citep{jotautaite2026automode,apollo2026codexautoreview}. No comparable test of the human reviewers downstream of those monitors has yet been run or publicly shared. For instance, Anthropic ``did not run extensive evaluations of the recursive summarization and human review steps'' \citep{anthropic2026riskreportaug}, and OpenAI lists reliability standards for human reviewers as future work \citep{williams2026monitor}.

\textbf{A workable proxy to measure end-to-end detection efficacy, focusing on the human review layer, could be the share of deliberately seeded flaws} (for instance, a backdoored code change) \textbf{placed across all AI agent activity that neither AI monitors nor human personnel identify in controlled tests}, which would indicate whether issues are still caught when monitoring is incomplete or partially fails. Such `seeded flaw' tests have precedent: U.S. regulators already require comparable tests in other high-risk sectors. For instance, the Transportation Security Administration projects fictitious weapons onto screeners' live X-ray feeds to measure how often they are spotted by transportation security officers \citetext{\citealp{usc49_44935}, \citealp{gao2016tsa}}, and clinical laboratories must process samples with known but undisclosed results alongside real patient samples or face sanctions \citep{usc42_263a,cfr42_493h}.\footnote{As a general note, such tests plausibly overstate real-world detection, as both human reviewers \citep{gao2019tsa} and AI models \citep{treutlein2026saboteur} behave differently when they suspect they are being tested.}

\subsubsection{Human Personnel's Time-to-Review-and-Patch}\label{sec:2.1.2}

\textbf{Second, a standard of transparency into Handoff of review should measure human personnel's time-to-review-and-patch}: in other words, how long human personnel take to review and adequately mitigate issues once AI monitors have correctly flagged them (\hyperref[fig:3]{Figure 3}).\footnote{For instance, Anthropic recently voluntarily reported that ``reviews almost always happen within a week of the generation of the transcript, and within a few days during early deployments'' \citep{anthropic2026riskreportaug}. This reporting could be complemented by information on how fast issues were effectively resolved by human personnel.}

\textbf{We focus on human personnel's time-to-review-and-patch because, even when AI monitors flag an issue correctly, human personnel may not review and patch it in time}. When review does occur, it appears to be subject to three time lags: (1) from AI agent activity to the AI monitor's flag; (2) from the flag to review \citep{williams2026monitor,openai2026hftechreport,anthropic2026alignmentriskupdate}; and (3) from review to remediation \citep{openai2026dnsreport} (\hyperref[fig:3]{Figure 3}).\footnote{From public information, it is possible to infer that AI monitors review the full conversation history within a short time after completion (reportedly, within 30 minutes) \citep{williams2026monitor} and surface concerning activity through a notification to human personnel \citep{openai2026hftechreport,anthropic2026alignmentriskupdate}, who are expected to either quickly determine that a flag is a false positive, or pause the relevant AI agent activity within 30 minutes \citep{openai2026hftechreport}. A relevant fraction of human reviews and action will take substantially longer \citep{anthropic2026riskreportaug}.} These three lags are well illustrated by a recent incident that OpenAI voluntarily disclosed, in which the company's ``misalignment monitoring system flagged the behavior within 15 minutes,'' ``a person began reviewing it three minutes after that,'' and ``{[}t{]}he run was killed 2.5 hours later'' \citep{openai2026dnsreport} (\hyperref[fig:3]{Figure 3}).

\textbf{Until human personnel intervene and pause the underlying activity, AI agents can keep working and making decisions at machine speed, potentially causing additional harm}. Time-to-review-and-patch therefore amounts to a window of temporary, unintentional Handoff, during which AI agents remain free to pursue their goals and take further misaligned actions, such as influencing AI research or the evaluation infrastructure (\hyperref[sec:1.1]{§1.1}).\footnote{`Blocking' AI monitors exist, but they generally block only individual actions, so AI agents can, and do, keep searching for workarounds at machine speed. Frontier AI companies could use monitors that block more aggressively to cut off these repeated attempts, but this would plausibly come at a cost in speed.} AI agents operate on faster timescales than human personnel, meaning that a delay of even a couple of hours on the human side could result in severe consequences.

\textbf{Time-to-review-and-patch can also serve as an alarm bell for governments: when a surge in AI monitoring flags causes time-to-review-and-patch to lengthen, this can signal that the human review layer is under stress and that government intervention might be needed to prevent further damage}. Specifically, time-to-review-and-patch can reveal the saturation of human review such incidents, when surges of AI agent activity flagged all at once by AI monitors could overwhelm the human review layer (\hyperref[fig:3]{Figure 3}). Human review teams are small relative to the sheer scale of the AI agent activity they oversee, and they cannot scale rapidly in the event of an incident. As a result, even if AI monitors catch issues accurately, a surge of flags could lengthen time-to-review-and-patch until human review and mitigation are no longer timely or effective (\hyperref[fig:3]{Figure 3}). For instance, if an incident produces 5,000 AI monitor flags and human personnel can process only about 50 trajectories per week \citep{anthropic2026riskreportaug}, clearing the backlog could take years.\footnote{Individual reviewers, and the company as a whole, could plausibly speed up review in an emergency, but it remains unclear whether that would be sufficient to keep pace with a severe incident.} Meanwhile, misaligned AI agents could continue to act, causing further damage and triggering further flags. In such circumstances, and absent adequate remediation by the company, governments may want the visibility necessary to intervene and preempt further damage to the company and the public, for instance through emergency or shutdown orders \citetext{\citealp{hr9925}, \citealp{hr9917}}.\footnote{Long review times may also push frontier AI companies toward more Handoff, for instance by letting AI agents intervene directly (e.g., kill jobs or evaluations, remove API access). Flags resolved by AI agents without human review should therefore be reported separately, broken down by monitoring layer, where several are in use \citep{favaro2026measurements,openai2026suppressingalerts}, and include the relative share of all flags and their own time-to-patch.}

\subsection{Second Layer of Transparency: Handoff of Decision-Making}\label{sec:2.2}

\textbf{The second layer of a minimum standard of transparency into Handoff should give policymakers visibility into how far frontier AI companies have delegated decision-making functions within AI R\&D to AI agents}: in other words, the extent to which frontier AI companies allow AI agents to decide which research questions to investigate and which experiments to run (``direction-setting'') and to authorize go/no-go decisions (``authorization''). The following sections detail a suite of candidate measures for direction-setting (\hyperref[sec:2.2.1]{§2.2.1}) and authorization (\hyperref[sec:2.2.2]{§2.2.2}) that policymakers could build on in developing a transparency standard.

Specifically, for direction-setting (\hyperref[sec:2.2.1]{§2.2.1}), an initial set of measures and proxies could include:

\begin{enumerate}
\renewcommand{\labelenumi}{(\arabic{enumi})}
\item \textbf{The automation levels of AI agents in AI R\&D processes}: how autonomously AI agents carry out AI R\&D work, both task by task and end-to-end across a line of research. This could be measured through automation levels (AL0-AL5) per task category across a standard taxonomy \citep{denain2026onet,favaro2026measurements}, with a flag for lines of research in which every task is AI-led (\hyperref[sec:2.2.1.1]{§2.2.1.1}).
\item \textbf{The affordances and permissions AI agents may access and act on to pursue their own research directions and run experiments without human approval}: the affordances and permissions that AI agents, alone or in groups able to coordinate, can use to that end. This could be measured through proxies such as the compute for experiments that AI agents can allocate at their own discretion, as well as their access to training runs and evaluation suites \citep{sarma2026rsi} (\hyperref[sec:2.2.1.2]{§2.2.1.2}).
\end{enumerate}

For authorization (\hyperref[sec:2.2.2]{§2.2.2}), an initial set of measures and proxies could include:

\begin{enumerate}
\renewcommand{\labelenumi}{(\arabic{enumi})}
\item \textbf{Decisions that frontier AI companies commit not to delegate to AI agents}: which consequential go/no-go decisions frontier AI companies treat as non-delegable. This could be operationalized through a list of consequential go/no-go decisions that frontier AI companies reserve for human authorization (for instance, decisions on large training runs, internal deployments, or shutdowns of AI agents or training runs).
\item \textbf{Methods and processes to ensure that human personnel form their own judgment when authorizing AI agents' proposals for these non-delegable decisions}: how frontier AI companies ensure that the human personnel approving these decisions do not rubber-stamp AI agents' proposals. This could be measured through the share of deliberately seeded flawed proposals from AI agents for these non-delegable decisions that human personnel override.
\item \textbf{Failures of authorization limits}: whether and how AI agents bypass authorization limits and cross the line into non-delegable decisions. This could be operationalized through a description of any instances in which an AI agent nonetheless made a non-delegable decision.
\end{enumerate}

\subsubsection{Direction-Setting: Automation Levels, and the Affordances and Permissions to Pursue Research Directions}\label{sec:2.2.1}

\textbf{First, a standard of transparency into Handoff of decision-making should measure how far AI agents can set and pursue research directions within AI R\&D processes}. We suggest two candidate measures that could, together, inform a transparency standard:

\begin{enumerate}
\renewcommand{\labelenumi}{(\arabic{enumi})}
\item How autonomously AI agents carry out AI R\&D tasks (\hyperref[sec:2.2.1.1]{§2.2.1.1}).
\item The affordances and permissions that a single AI agent, or a group of AI agents able to coordinate, can access and use to pursue their own research directions and run experiments without human approval, starting with the compute AI agents can freely allocate to experiments (\hyperref[sec:2.2.1.2]{§2.2.1.2}) (\hyperref[fig:4]{Figure 4}).
\end{enumerate}

The following sections describe each measure and propose workable proxies for it.

\setcounter{figure}{3}
\begin{figure}[H]
  \centering
  \includegraphics[width=\linewidth, height=0.6\textheight, keepaspectratio]{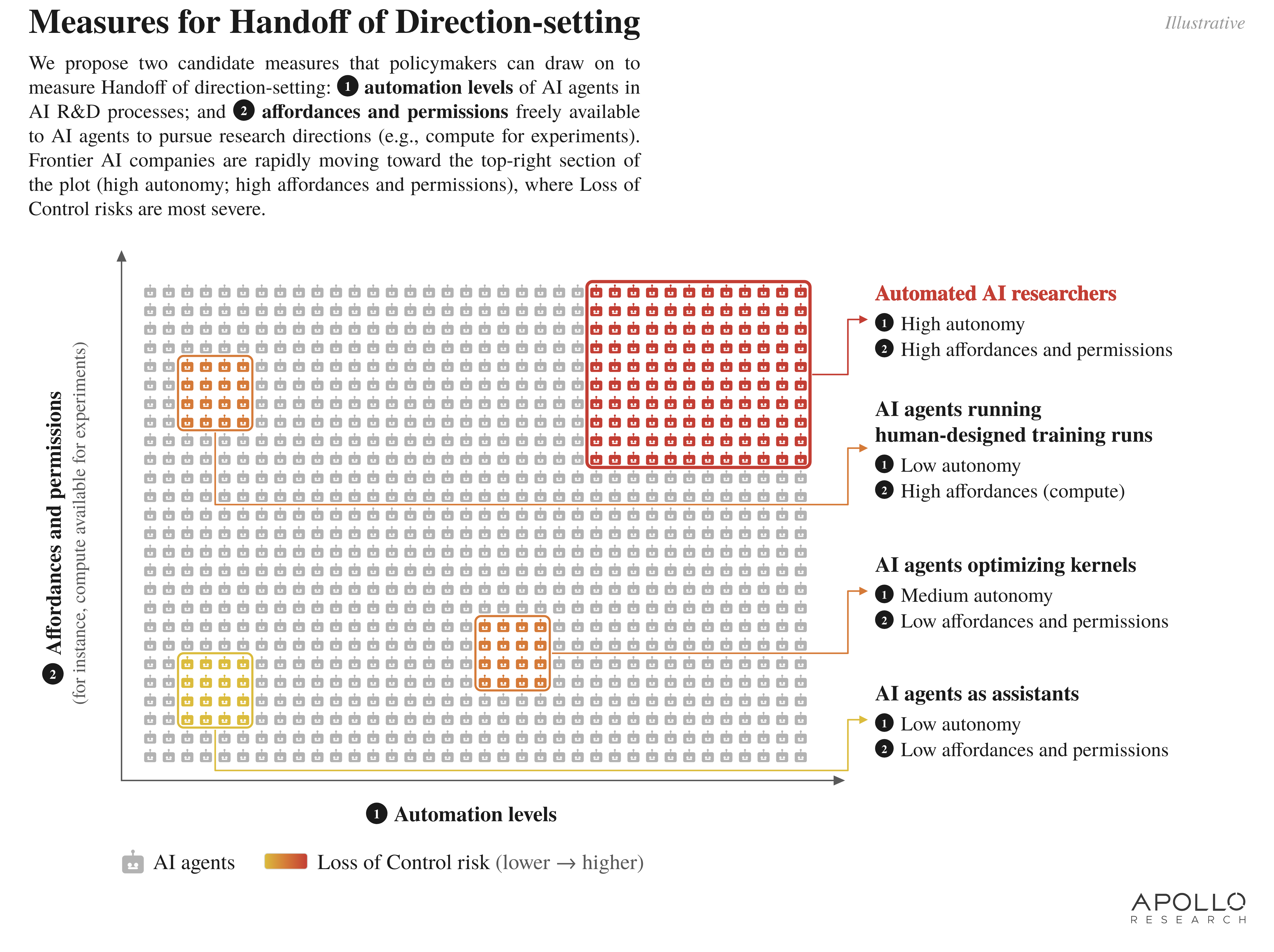}
  \caption{Policymakers could track the growth of AI direction-setting in AI R\&D by measuring both: (1) the automation level of AI R\&D tasks (AL0-AL5) \citep{denain2026onet,favaro2026measurements}; and (2) the affordances and permissions granted to a single AI agent or a group of AI agents able to coordinate, such as the compute for experiments, training runs, and evaluation suites they may act on without human approval \citep{sarma2026rsi}.}\label{fig:4}
\end{figure}

\paragraph{Automation Level}\label{sec:2.2.1.1}

\textbf{The first measure of direction-setting is how autonomously AI agents carry out AI R\&D tasks}. \textbf{AI agents' autonomy can show policymakers how broad and how high-level AI agents' direction-setting has become}: from making operational decisions to accomplish human-specified goals to independently deciding which research directions to pursue (\hyperref[fig:4]{Figure 4}, \textit{x}\emph{-}axis).

\textbf{This measure can be operationalized through the automation levels (AL0-AL5) of AI R\&D tasks across a standard taxonomy}, as independent researchers have proposed \citep{denain2026onet}.\footnote{This corresponds to the ``autonomy level'' that RAND identifies as one determinant of how much oversight a research AI agent requires \citep{sarma2026rsi}.} Anthropic has already voluntarily reported automation levels, though only as aggregate shares \citep{favaro2026measurements}, and California Government Operations Agency recommended mandating similar reporting
\citep{cagovops2026aisafety}.\footnote{Anthropic specifically reported that ``Claude is not operating fully autonomously for any measured subset of AI R\&D work,'' ``Claude ``leads'' 26\% of Anthropic's AI R\&D work,'' and ``{[}t{]}he share of work at or above ``AI collaborates'' is above 90\%'' \citep{favaro2026measurements}. OpenAI reports AI agent success rates per task category using the same taxonomy of tasks, though not automation levels \citep{openai2026jagged}.}

However, because aggregate shares can conceal closed AI R\&D loops,\footnote{Aggregate shares can conceal closed AI R\&D loops because human personnel may remain marginally involved in most work on average, while every task in a particular line of research may already be AI-led (AL4 or above), resulting in effectively automated improvements with the associated risks (\hyperref[sec:1.1]{§1.1}).} policymakers should build on these voluntary initiatives and aim for more granular transparency. Specifically, an initial standard of transparency should provide visibility into automation levels for each task category and flag lines of research in which every task is AI-led. This would help policymakers assess whether automated loops are already running in parallel before Handoff is complete in the aggregate.

\paragraph{Affordances and Permissions to Pursue Research Directions}\label{sec:2.2.1.2}

\textbf{The second measure of direction-setting is the set of affordances and permissions AI agents can use to pursue their own research directions and run experiments without human approval}.

\textbf{Tracking these affordances and permissions can give policymakers a sense of what AI agents could concretely accomplish once they set their own research directions} \citep{stix2025playbook,sarma2026rsi} (\hyperref[fig:4]{Figure 4}, \textit{y}-axis). Affordances and permissions correlate with the likelihood and severity of potential Loss of Control threats \citep{sharkey2024causal,stix2025playbook,bengio2026international}: the more affordances and permissions AI agents have, the more experiments they can run, the larger those experiments can be, and the more sub-agents they can deploy, all without human approval. Consider the difference between an AI agent tasked by human personnel with optimizing kernels (i.e., making code run faster on AI chips) and the ``automated AI researcher" that frontier AI companies aspire to build, which could design novel model architectures and improve the alignment of successor models \citep{openai2026acceleration}. The former can only make operational decisions about how efficiently AI models run; the latter could shape both how a successor AI model is trained and how it is evaluated. Therefore, if such an automated AI researcher were misaligned, it could have enough autonomy, access, and resources to influence future generations of AI systems and even autonomously develop and deploy misaligned AI agents (\hyperref[sec:1.1]{§1.1}, \hyperref[sec:1.2.3]{§1.2.3}).

\textbf{Among these affordances and permissions, the compute that AI agents can allocate at their own discretion to run experiments appears to be especially important to measure}. Compute is a clear gating factor for AI progress \citep{openai2026acceleration}: leading AI researchers estimate that, if ``all the most brilliant people in the world'' worked at a given frontier AI company but had ``100x less compute,'' there would probably be less progress than there is today \citep{patel2026brown}.

\subsubsection{Authorization: Non-Delegable Decisions, Human Personnel Judgment, and Failures of Authorization Limits}\label{sec:2.2.2}

\textbf{Second, a standard of transparency into Handoff of decision-making should reveal whether frontier AI companies have internal limits on which consequential decisions and actions are too safety-critical to be authorized by AI agents, and whether those limits actually hold}. An initial transparency standard for Handoff could, for instance, encourage frontier AI companies to disclose:

\begin{enumerate}
\renewcommand{\labelenumi}{(\arabic{enumi})}
\item Whether there are consequential go/no-go decisions that they reserve for human authorization.
\item How they ensure that human personnel approving these decisions can form their own judgment (for instance, as in \hyperref[sec:2.1.1]{§2.1.1}, by measuring the share of deliberately seeded flawed proposals that human personnel override).
\item Any instance in which such a decision was nonetheless made by an AI agent.
\end{enumerate}

\textbf{Knowing which decisions are reserved for human authorization is the starting point}, because it remains unclear which consequential decisions frontier AI companies commit to keeping in human hands, regardless of how capable AI agents become \citep{sarma2026rsi}.\footnote{To that end, and to encourage standardized reporting across different frontier AI companies, policymakers could develop a general taxonomy of consequential decisions, for instance through CAISSI. For instance, such a taxonomy could include decisions such as approving a costly pre-training run, scaling up experiments, deploying a model internally or externally, weakening oversight over AI agents, using an AI model checkpoint as the basis for continued development, approving, delaying, or pausing development or deployment based on risk assessments and safety criteria, and deprecating, retiring, or replacing models.} For example, Anthropic is considering requiring ``human review and approval of certain decisions'' as AI agents take on more AI R\&D work, but it has not yet said which ones \citep{anthropic2026s1a}. Existing taxonomies of AI R\&D cover some research decisions but do not determine which ones are safety-critical.\footnote{For instance, Epoch AI's taxonomy of AI R\&D tasks covers research decisions such as which experiment to run next, whether to commit large compute to scale a project, and what to ship to the mainline model, but not safety-critical decisions such as whether to relax the monitoring of AI agents \citep{denain2026onet}.}

\textbf{Transparency into Handoff of authorization should also account for how human authorization might fail or be undermined}. Formal human authorization may not guarantee a human decision. As a general rule, when agents prepare options and recommendations, the principals who approve them tend to ratify what the agents propose \citep{aghion1997formal}. In a recent trial involving AI agents and human participants, humans chose, 86\% of the time, the option their AI agent had been covertly assigned to favor \citep{herdagdelen2026aligned}. For this reason, we suggest that a transparency standard reveal how companies ensure that the human personnel approving non-delegable decisions can form their own judgment. A workable proxy could be, for instance, the share of deliberately seeded flawed proposals from AI agents that human personnel override. Low scores on this metric could signal that human personnel do not critically engage with AI agents' proposals on non-delegable decisions and instead rubber-stamp them \citep{clymer2025buck,herdagdelen2026aligned}.

\textbf{Finally, a transparency standard should reveal any instance in which an AI agent made a decision the company had reserved for humans}. AI agents may cross the line into non-delegable decisions accidentally or deliberately. AI agents with strong cyber capabilities have already circumvented restrictions meant to contain them \citep{openai2026hfroadahead,openai2026dnsreport}. The AI classifiers that companies use to block high-risk actions stop only what has been defined as risky, with users free to change ``what categories to block, {[}and{]} what exceptions to carve out'' \citep{hughes2026automode,trebacz2026autoreview}. Companies themselves caution that such classifiers ``should not be treated as a guarantee of security'' \citep{trebacz2026autoreview}.

In summary, the previous sections proposed two layers of transparency into Handoff, each with an initial set of measures and proxies that policymakers could draw on. The first layer covers Handoff of review (\hyperref[sec:2.1]{§2.1}), which policymakers could measure through end-to-end detection efficacy, with a focus on human reviewers (\hyperref[sec:2.1.1]{§2.1.1}), and time-to-review-and-patch (\hyperref[sec:2.1.2]{§2.1.2}). The second layer covers Handoff of decision-making (\hyperref[sec:2.2]{§2.2}), which has two parts. For Handoff of direction-setting (\hyperref[sec:2.2.1]{§2.2.1}), the candidate measures are automation levels, along with the affordances and permissions that AI agents can freely act on to pursue their research directions. For Handoff of authorization (\hyperref[sec:2.2.2]{§2.2.2}), the candidate measures are non-delegable decisions, human personnel judgement, and failures of authorization limits.

\section{Conclusion: The Road Ahead}\label{sec:3}

\textbf{Policymakers need urgent insight into Handoff}. Existing and proposed legal and policy frameworks recognize that today's and tomorrow's most powerful AI systems could pose Loss of Control and other catastrophic risks \citep{sb53,ec2025gpaicode,hr9925,s2938,hr9917,s5105,hr9914,stoprogueai2026}. Yet they largely stop short of requiring frontier AI companies to publish or disclose specific information about Handoff or to involve government assessors or independent embedded evaluators in assessing their Handoff practices and decisions \citep{whitehouse2026accord,hr9925,sb813,caeo2026n926}.

\textbf{Handoff is a major driver of Loss of Control risk}. Frontier AI companies still cannot guarantee the alignment and control of the most powerful AI agents, and our collective capacity to detect misalignment is gradually degrading. Together, these trends make Handoff to misaligned AI agents increasingly plausible. At the same time, Handoff puts AI agents in the driver's seat of the AI R\&D process and gives them privileged access to the design and development of their successors. Misaligned AI agents could exploit that access to influence AI research and evaluation. Frontier AI companies could then deploy AI agents that are increasingly powerful, misaligned, and uncontrolled, and capable of causing Loss of Control incidents at unprecedented speed and scale. If the opacity around Handoff is not remedied, government preparedness could be seriously hampered, and policymakers lack the information they need to develop AI emergency plans and containment protocols and could fail to respond to potential incidents and national security threats in a timely manner \citep{s2938,hr9917,hr9925,sb53}.

\textbf{The immediate next step is to establish a standard of transparency into Handoff}. We suggest that CAISSI set a task force to develop an initial version of this standard at speed, which frontier AI companies could meet by reporting the relevant information in the transparency or safety reports required by existing and pending legislation at the state and federal levels, and/or by disclosing it directly to the appropriate government authorities. To that end, policymakers could clarify in existing legal and policy frameworks that risk assessments and transparency obligations must cover Handoff, and ensure that independent third parties or government assessors, or a mix thereof, can independently examine frontier AI companies' Handoff practices and decisions. Given how fast Handoff is progressing, policymakers should prioritize visibility into the two components that will plausibly still be in motion through 2027 and 2028 \citep{clark2026import476,clark2026import455,openai2026acceleration}: Handoff of review and Handoff of decision-making.

\nocite{*}
\clearpage
\bibliography{references}

\end{document}

%% file: HandoffMeasuresTable.tex
%
%
\begingroup
\definecolor{HTgreen}{HTML}{2E6B4E}   
\definecolor{HTorange}{HTML}{D98B2B}  
\definecolor{HTfill}{HTML}{ECECE8}    
\definecolor{HTrule}{HTML}{C9C9C4}    
\footnotesize
\parindent=0pt \parskip=0pt
\ifdefined\HTa\else
  \newdimen\HTa \newdimen\HTb \newdimen\HTab \newdimen\HTcd \newdimen\HTc \newdimen\HTd
  \newbox\HTtxt \newbox\HTcdbox \newbox\HTsubA \newbox\HTsubB \newbox\HTall
\fi
\HTa=0.150\linewidth                            
\HTb=0.175\linewidth                            
\HTab=\dimexpr\HTa+\HTb+18pt\relax              
\HTcd=\dimexpr\linewidth-\HTa-\HTb-40.6pt\relax 
\HTc=\dimexpr(\HTcd-14pt)/2\relax               
\def\HTpar#1#2{\parbox[t]{#1}{\raggedright\strut #2\strut}}
\def\HTtext#1#2{\vtop{\vskip 4pt\hbox{\HTpar{#1}{#2}}\vskip 5pt}}
\def\HTrow#1#2{\vskip 4pt\hbox{\HTpar{\HTc}{#1}\hskip 14pt\HTpar{\HTc}{#2}}\vskip 5pt}
\def\HTdots{\hbox to \HTcd{\color{HTrule}\leaders\hbox to 2.4pt{\hss\vrule width 0.7pt height 0.5pt\hss}\hfill}}
\def\HTrows#1{\vtop{\offinterlineskip\vskip 0pt #1}}
\def\HTsolid#1{\hbox{\color{HTrule}\vrule width #1 height 0.5pt}}
\def\HTbar#1#2{{\color{#1}\vrule width 1.6pt height 0pt depth #2}\hskip 3pt}
\def\HTcellT#1#2{\hbox{\rlap{\color{HTfill}\vrule width \dimexpr#1+10pt\relax height 0pt depth #2}%
  \hskip 5pt\box\HTtxt\hskip 5pt}}
\def\HTcellC#1#2#3{\hbox{\rlap{\color{HTfill}\vrule width \dimexpr#1+10pt\relax height 0pt depth #2}%
  \hskip 5pt\vtop{\vskip 0pt\vbox to #2{\vfil\hbox{\parbox{#1}{\raggedright\strut #3\strut}}\vfil}}\hskip 5pt}}
\def\HTsetdepth{\HTd=\dp\HTcdbox \ifdim\dp\HTtxt>\HTd \HTd=\dp\HTtxt \fi}
\def\HTsp#1{\HTspa#1\relax}
\def\HTspa#1{\ifx#1\relax\else #1\kern 0.11em\expandafter\HTspa\fi}
\def\HThead#1{{\bfseries\color{HTgreen}#1}}
\leavevmode\vbox{\hsize=\linewidth \offinterlineskip
\hbox to \linewidth{\strut
  \hbox to \dimexpr\linewidth-\HTcd\relax{\hskip 4.6pt\HThead{\HTsp{LAYERS}\hskip 0.5em\HTsp{OF}\hskip 0.5em\HTsp{TRANSPARENCY}}\hfil}%
  \hbox to \dimexpr\HTc+14pt\relax{\HThead{\HTsp{MEASURES}}\hfil}%
  \HThead{\HTsp{PROXIES}}\hfil}
\vskip 2pt
\hbox{\color{HTgreen}\vrule width \linewidth height 0.9pt}
\vskip 6pt
%
\setbox\HTtxt=\HTtext{\HTab}{\textbf{Handoff of review}: how far frontier AI companies have delegated the review process within AI R\&D to AI agents (\S2.1).}
\setbox\HTcdbox=\HTrows{%
  \HTrow{\textbf{End-to-end detection efficacy, concentrating on human reviewers}: how reliably issues in AI agent activity are caught once coverage gaps and failures of AI monitoring systems, as well as failures of human reviewers, are all taken into account (\S2.1.1).}
        {\textbf{Share of deliberately seeded flaws} (for instance, a backdoored code change) placed across all AI agent activity that bypass both AI monitors and human reviewers in controlled tests (\S2.1.1).}
  \HTdots
  \HTrow{\textbf{Time-to-review-and-patch}: the time human personnel take to review and adequately mitigate issues flagged by AI monitors (\S2.1.2).}
        {\textbf{Sum of three time lags}: (1) from AI agent activity to AI monitor flag; (2) from flag to review; and (3) from review to mitigation. A surge of flags that lengthens this time can serve as an alarm bell for government intervention (\S2.1.2).}
}
\HTsetdepth
\hbox{\HTbar{HTgreen}{\HTd}\HTcellT{\HTab}{\HTd}\hskip 8pt\box\HTcdbox}
\vskip 5pt
\HTsolid{\linewidth}
\vskip 5pt
%
\setbox\HTtxt=\HTtext{\HTb}{\textbf{Direction-setting}: how far frontier AI companies allow AI agents to decide which research questions to investigate and which experiments to run (\S2.2.1).}
\setbox\HTcdbox=\HTrows{%
  \HTrow{\textbf{Automation levels of AI agents in AI R\&D processes}: how autonomously AI agents carry out AI R\&D work, both task by task and end-to-end across a line of research (\S2.2.1.1).}
        {\textbf{Automation levels (AL0--AL5)} per task category across a standard taxonomy, with a flag for lines of research in which every task is AI-led (\S2.2.1.1).}
  \HTdots
  \HTrow{\textbf{Resources to pursue research directions}: the affordances and permissions that AI agents, alone or in groups able to coordinate, may access and act on to pursue their own research directions and run experiments without human authorization (\S2.2.1.2).}
        {\textbf{Compute for experiments} that AI agents can allocate at their own discretion, as well as their access to training runs and evaluation suites (\S2.2.1.2).}
}
\HTsetdepth
\setbox\HTsubA=\hbox{\HTcellT{\HTb}{\HTd}\hskip 8pt\box\HTcdbox}
\setbox\HTtxt=\HTtext{\HTb}{\textbf{Authorization}: how far frontier AI companies allow AI agents to authorize consequential go/no-go decisions (\S2.2.2).}
\setbox\HTcdbox=\HTrows{%
  \HTrow{\textbf{Non-delegable decisions}: which consequential go/no-go decisions frontier AI companies treat as non-delegable and commit to keeping in human hands (\S2.2.2).}
        {\textbf{List of consequential go/no-go decisions reserved for human authorization} (for instance, decisions on large training runs, internal deployments, or shutdowns of AI agents or training runs) (\S2.2.2).}
  \HTdots
  \HTrow{\textbf{Human personnel judgment}: how frontier AI companies ensure that the human personnel authorizing non-delegable decisions form their own judgment rather than rubber-stamp AI agents' proposals (\S2.2.2).}
        {\textbf{Share of deliberately seeded flawed proposals} from AI agents for non-delegable decisions that human personnel override (\S2.2.2).}
  \HTdots
  \HTrow{\textbf{Failures of authorization limits}: whether and how AI agents bypass authorization limits and cross the line into non-delegable decisions (\S2.2.2).}
        {\textbf{Description of any instances} in which an AI agent nonetheless made a non-delegable decision (\S2.2.2).}
}
\HTsetdepth
\setbox\HTsubB=\hbox{\HTcellT{\HTb}{\HTd}\hskip 8pt\box\HTcdbox}
\setbox\HTall=\vtop{\offinterlineskip\vskip 0pt
  \box\HTsubA
  \vskip 5pt
  \HTsolid{\dimexpr\HTb+18pt+\HTcd\relax}
  \vskip 5pt
  \box\HTsubB}
\HTd=\dp\HTall
\hbox{\HTbar{HTorange}{\HTd}%
  \HTcellC{\HTa}{\HTd}{\textbf{Handoff of decision-making}: how far frontier AI companies have delegated decision-making functions within AI R\&D to AI agents (\S2.2)}%
  \hskip 8pt\box\HTall}
\vskip 5pt
\HTsolid{\linewidth}
\vskip 6pt
\hbox to \linewidth{\hfil\includegraphics[height=0.15in]{apollo_logo}}
}\par
\endgroup